\documentclass[12pt,aps,prb,preprint,showpacs,showkeys]{revtex4}  

\usepackage{amsmath}    
\usepackage{amsfonts}   
\usepackage{amssymb}
\usepackage{graphicx}   
\usepackage{subfigure}

\begin{document}

\title{Phase transitions in a triangular Blume-Capel antiferromagnet}
\author{M. \v{Z}ukovi\v{c}}
 \email{milan.zukovic@upjs.sk}
\author{A. Bob\'ak}
 \affiliation{Department of Theoretical Physics and Astrophysics, Faculty of Science,\\ 
P. J. \v{S}af\'arik University, Park Angelinum 9, 041 54 Ko\v{s}ice, Slovakia}
\date{\today}

\begin{abstract}
We study the critical behavior of a frustrated Blume-Capel (BC) antiferromagnet on a triangular lattice by Monte Carlo simulations. For a reduced single-ion anisotropy strength $-1.47 \lesssim d < 0$ we find two phase transitions. The low-temperature phase is characterized by the antiferromagnetic long-range ordering (LRO) on two sublattices with the third one remaining in a non-magnetic state. At higher temperatures there is a critical region of the Berezinskii-Kosterlitz-Thouless (BKT) type with a power-law decaying spin-correlation function. For $-1.5 \le d \lesssim -1.47$, there is only one phase transition from the LRO to the paramagnetic region and the transition is of first order. The presence of the BKT phase in the current frustrated BC model is a new feature not observed in its non-frustrated counterparts. The values of the decay exponent $\eta$ of the BKT phase corresponding to upper and lower temperatures appear to be consistent with the theoretical predictions for the six-state clock model.
\end{abstract}



\maketitle

\section{INTRODUCTION}

It has been shown exactly that in the fully frustrated triangular lattice Ising antiferromagnet (TLIA) with spin 1/2 no long-range order can exist down to zero temperature~\cite{wann} but the ground state is critical with the power-law decaying spin-correlation function~\cite{step}. However, the situation can change dramatically for larger spin values. The series of studies~\cite{naga,yama,lipo,zeng} have argued that long-range order (LRO) can occur in the ground state if the spin is larger than some critical value. The corresponding spin structure is of the type $(1,-1,0)$, i.e., with two sublattices of opposite magnetizations and one sublattice of zero magnetization. The upper bound of this critical value was estimated by the use of Peierls' argument~\cite{naga} as 62 and a more precise value was established by Monte Carlo simulations~\cite{yama} as 11/2. Generally, the lack of order in frustrated spin systems is due to large ground-state degeneracy and the above studies have shown that such degeneracy can be considerably affected by the spin magnitude, which can lead to long-range ordering. Nevertheless, the large degeneracy can also be lifted by some other perturbations, resulting in long-range ordering even in the highly frustrated spin-1/2 system, such as an external magnetic field~\cite{metcalf,schick,netz,zuko1}, selective dilution~\cite{kaya,zuko2} or inclusion of the exchange interactions with further neighbors~\cite{metcalf1,tana,land,fuji,taka,rast}. 

It is well known that in the Ising models with spin larger than 1/2 a single-ion anisotropy is another parameter that may play a crucial role in their critical properties (see, e.g.,~\cite{cape,blum}). This so called Blume-Capel (BC) model has been intensively studied~\cite{blum1,graf,laze,jain,saul,berk,burk,siqu} mostly on bipartite lattices, in which case the sign of the exchange interaction is irrelevant to their critical properties in the absence of an external field. The model has been confirmed to belong to the standard Ising universality class~\cite{fyta}. However, for an antiferromagnetic BC model on non-bipartite lattices we can expect qualitatively different behavior. A frustrated antiferromagnetic spin-1 BC model on a triangular lattice has been investigated by position-space renormalization group (PSRG)~\cite{maha} and transfer matrix~\cite{coll} methods, and has been found to display a finite-temperature antiferromagnetic (AF) LRO of the type $(1,-1,0)$ within a certain range of the single-ion anisotropy strength, accompanied with a multicritical behavior. Nevertheless, 
the universality class of the identified second-order phase transition was not examined. On the other hand, it is known that a number of frustrated systems violate the ordinary universality hypothesis. For example, the spin-1/2 Ising antiferromagnet with the frustration arising from the competing nearest-neighbor (NN) and next-nearest-neighbor (NNN) interactions on a square lattice lead to a nonuniversal (or weakly universal) critical behavior in which the critical indices of the model depended on the NNN to NN interaction ratio~\cite{swen,bind,land1,tana1,mina}. Similar behavior was also found in the spin-1 model involving either the competing NN and NNN interactions~\cite{hond} or positive biquadratic interactions~\cite{bade}. 

Therefore, the motivation for the present investigations was to study the character of the critical behavior of the geometrically frustrated spin-1 antiferromagnet on a triangular lattice with the single-ion anisotropy by Monte Carlo simulations. Surprisingly, we found that the phase transition from the LRO phase is not second order to a paramagnetic phase, as predicted by the PSRG results~\cite{maha}, but of a Berezinskii-Kosterlitz-Thouless (BKT) type to a quasi-ordered phase with algebraically decaying spin correlation function, which persists for a range of intermediate temperatures between the LRO and paramagnetic phases. This finding is unexpected not only because the BKT phase was not found in the earlier investigations~\cite{maha,coll} but also because no BKT phase was predicted to exist at finite temperatures in the spin-1 TLIA model~\cite{lipo,zeng}.

\section{MODEL AND SIMULATION DETAILS}

We consider the model described by the Hamiltonian 
\begin{equation}
\label{Hamiltonian}
H=-J\sum_{\langle i,j \rangle}S_{i}S_{j}-D\sum_{i}S_{i}^2,
\end{equation}
where $S_{i}=\pm1,0$ is an Ising spin on the $i$th lattice site, $\langle i,j \rangle$ denotes the sum over nearest neighbors, $J<0$ is an antiferromagnetic exchange interaction parameter, and $D$ is a single-ion anisotropy parameter. 

In order to study phase transitions in the present spin system we employ Monte Carlo (MC) method. We perform MC simulations on spin systems of the size $L^2$, where $L=24,48,72,96$, and $120$. We apply the periodic boundary conditions and the updating follows the Metropolis dynamics. To obtain dependencies of various thermodynamic quantities on the reduced temperature $k_BT/|J|$, we use standard MC simulation in which for thermal averaging we typically consider up to $N=2 \times 10^6$ MCS (Monte Carlo sweeps or steps per spin) after discarding another $N_{0} = 0.2 \times N$ MCS for thermalization. The simulations start from high temperatures, using random initial configurations. Then the temperature is gradually lowered with the steps $k_B\Delta T/|J|=0.02$ (or $0.01$ around the critical region) and the simulations start from the final configuration obtained at the previous temperature. In order to obtain critical indices, we perform finite-size scaling (FSS) analysis, in which case we apply the reweighing techniques \cite{ferr88,ferr89} and use $N= 10^7$ MCS. We note that sufficiently long simulation times are necessary for the present system, since the integrated autocorrelation time at the criticality ranged from $\tau \sim 10^2$ MCS for $L=24$ up to $\tau \sim 10^3$ MCS for $L=120$, following the scaling law $\tau \propto L^z$ with the estimated exponent $z \approx 2.2$. For more reliable estimation of statistical errors, we used the $\Gamma$-method~\cite{wolf04}.

For an antiferromagnet, as an order parameter it is useful to define the staggered magnetization per site as
\begin{equation}
\label{mag_st}
m_s = \langle M_s \rangle/L^2 = 3\Big\langle \max\Big(\sum_{i \in A}S_{i}, \sum_{j \in B}S_{j}, \sum_{k \in C}S_{k}\Big)-\min\Big(\sum_{i \in A}S_{i}, \sum_{j \in B}S_{j}, \sum_{k \in C}S_{k}\Big) \Big\rangle/2L^2,
\end{equation}
\noindent where $\langle\cdots\rangle$ denotes the thermal average. Further, the following quantities which are functions of $H$ or/and $M_s$ are defined: 
the specific heat per site
\begin{equation}
\label{eq.c}c=\frac{\langle H^{2} \rangle - \langle H \rangle^{2}}{L^2k_{B}T^{2}},
\end{equation}
the staggered susceptibility per site
\begin{equation}
\label{eq.chi}\chi_{s} = \frac{\langle M_{s}^{2} \rangle - \langle M_{s} \rangle^{2}}{L^2k_{B}T},
\end{equation}
the derivatives of the following functions of $\langle M_s \rangle$ with respect to
$\beta=1/k_{B}T$
\begin{equation}
\label{eq.D1}D_{1s} = \frac{\partial}{\partial \beta}\ln\langle M_s \rangle = \frac{\langle M_s H
\rangle}{\langle M_s \rangle}- \langle H \rangle,
\end{equation}
\begin{equation}
\label{eq.D2}D_{2s} = \frac{\partial}{\partial \beta}\ln\langle M_{s}^{2} \rangle = \frac{\langle M_{s}^{2} H
\rangle}{\langle M_{s}^{2} \rangle}- \langle H \rangle,
\end{equation}
and the Binder parameter (magnetic fourth-order cumulant)
\begin{equation}
\label{eq.U}U = 1-\frac{\langle M_{s}^{4}\rangle}{3\langle M_{s}^{2}\rangle^{2}}.
\end{equation}
\hspace*{5mm} The above quantities are useful for localization of the phase boundaries as well as for determination of the
nature of the phase transition. For example, temperature-dependences of a variety of thermodynamic quantities display extrema at the $L$-dependent pseudo-transition temperatures $k_BT_c(L)/|J|$. Thus, for the second-order transition, the critical temperature can be estimated from the locations of the peaks of the response functions, such as $c$ and $\chi_s$, for a given value of $L$.
Then, the observed extrema are known to scale with a lattice size as, for example:
\begin{equation}
\label{eq.scalchi}\chi_{s}(L) \propto L^{\gamma/\nu},
\end{equation}
\begin{equation}
\label{eq.scalD1}D_{1s}(L) \propto L^{1/\nu},
\end{equation}
\begin{equation}
\label{eq.scalD2}D_{2s}(L) \propto L^{1/\nu},
\end{equation}
\noindent where $\gamma$ and $\nu$ represent the critical exponents of the staggered susceptibility and correlation length, respectively. More precise locations of the extrema used in FSS can be obtained by reweighing techniques applied to the simulation results performed at the pseudo-critical temperature $k_BT_c(L)/|J|$~\cite{ferr88,ferr89}.\\
\hspace*{5mm} Furthermore, it is known that in the ground state the sublattice spin-correlation function of the TLIA model decays as a power law~\cite{step}:
\begin{equation}
\label{PL}\langle S_{i}S_{j} \rangle \propto r_{ij}^{-\eta},
\end{equation}
where $\eta$ is the critical exponent of the correlation function. The exponent $\eta$ of the model with zero single-ion anisotropy has been shown to decrease with the spin value from $\eta=1/2$ for spin-1/2 to zero for spin larger than 11/2, for which the AF LRO occurs~\cite{naga,yama}. Power-law decay of the spin-correlation function is a characteristic of the Berezinskii-Kosterlitz-Thouless (BKT) phase~\cite{kost} and the exponent $\eta$ can be estimated by FSS of the order parameter $m_s$, which scales as~\cite{chal}
\begin{equation}
\label{ms_FSS}
m_s(L) \propto L^{-\eta/2}.
\end{equation}
Alternatively, it can also be obtained from the staggered susceptibility, which in the BKT phase where $\langle M_s \rangle$ vanishes in the infinite lattice size limit is more appropriately defined as~\cite{chal,rast1}
\begin{equation}
\label{eq.chi_bkt}\chi_{s}' = \frac{\langle M_{s}^{2} \rangle }{L^2k_{B}T},
\end{equation}
and which scales as
\begin{equation}
\label{xis_FSS}
\chi_{s}'(L) \propto L^{2-\eta}.
\end{equation}
In order to distinguish between the second-order and the BKT transitions, one can employ a so called cumulant method~\cite{bin,chal}, which is based on the behavior of the Binder parameter $U$ using the formula
\begin{equation}
\label{U_FSS}
\partial (U(L')/\partial U(L))_{T_c} = (L'/L)^{1/\nu}.
\end{equation}
In the case of a second-order transition at the critical temperature $T_c$ the exponent $\nu$ is finite and the correlation length diverges as $[(T_c-T)/T_c]^{-\nu}$. On the other hand, in the case of a BKT transition $\nu \rightarrow \infty$, singularities are exponential and the correlation length diverges as  
\begin{equation}
\label{xi_bkt}
\xi = \xi_0 \exp(a[(T_c-T)/T_c]^{-1/2}).
\end{equation}
Then, if the BKT phase is expected between the long-range ordered (LRO) and the paramagnetic (P) phases, the following scaling relations apply:
\begin{equation}
\label{ms_bkt}
m_sL^{b} = f_1(L^{-1}\exp(at^{-1/2})),
\end{equation}
where $b=\eta/2$, $t=(T_1-T)/T_1$, $T<T_1$, and $T_1$ is the LRO-BKT transition temperature, and
\begin{equation}
\label{xis_bkt}
\chi_s'L^{-c} = f_2(L^{-1}\exp(at^{-1/2})),
\end{equation}
where $c=2-\eta$, $t=(T-T_2)/T_2$, $T>T_2$, and $T_2$ is the BKT-P transition temperature.

\section{RESULTS AND DISCUSSION}
\begin{figure}[b]
\centering
    \includegraphics[scale=0.5]{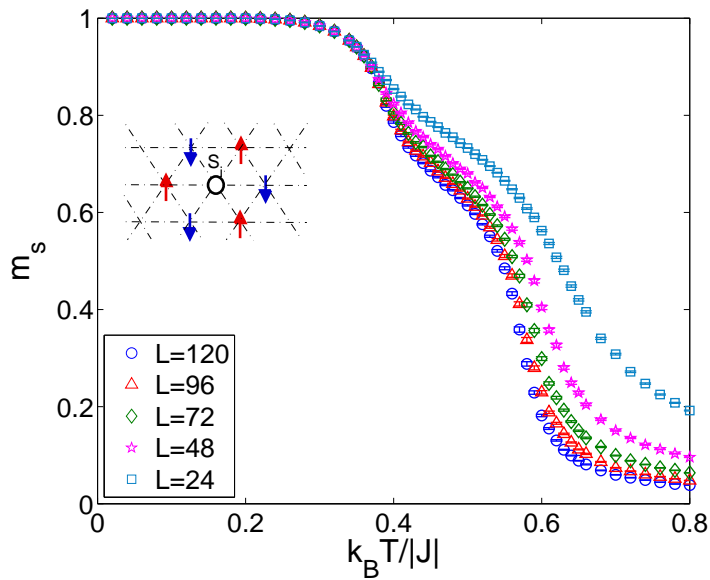}
\caption{(Color online) Temperature variation of the staggered magnetization $m_s$, for $d=-1$ and different values of $L$. The inset shows the spin arrangement on a hexagonal plaquette with the ``free'' central spin $S_i$. The up and down arrows denote the spin values $+1$ and $-1$, respectively.}\label{fig:ms-T}
\end{figure}

Let us first examine the ground state properties for different values of $d \equiv D/|J|$. For $d=0$ the ground-state configuration is such that the spins on each elementary triangular plaquette sum to $\pm 1$. Hence, if we consider a hexagonal plaquette with the antiferromagnetic arrangement of the nearest neighbors on the honeycomb backbone, as shown in the inset of Fig.~\ref{fig:ms-T}, the central $S_{i}$ spin is ``free'' and the configurations with any value of $S_{i}=\pm1,0$ are energetically equivalent. If $d>0$ the configurations with the values of $S_{i}=\pm1$ are preferred and the system behaves like a spin-1/2 Ising model with no long-range order~\cite{wann}. If $d<0$ the configurations with the values of the central spin $S_{i}=\pm1$ are suppressed and the ordered phase with the antiferromagnetic ordering on the honeycomb backbone and non-magnetic states of the central spins, i.e., $S_{i}=0$, can occur. The triangular patterns of the type ($1,-1,0$) are six-fold degenerate. However, such a state is only favorable for $-3/2<d<0$. Below $d=-3/2$ the energy becomes positive and therefore the non-magnetic state with all the spins taking zero value is the ground state. The above cases are summarized in Table~\ref{tab:GS}.

\begin{table}[t]
\caption{Types of ground-state spin configurations on triangular plaquettes with the corresponding energies per site $\langle H \rangle/|J|N$, for different anisotropy $d$ intervals.}
\label{tab:GS}
\centering
\begin{tabular}{lccc}
\hline
$d$  & $\left(-\infty,-3/2\right)$  & $\left(-3/2,0\right)$ & $(0,\infty)$ \\
\hline
($S_1,S_2,S_3$) & ($0,0,0$)  & ($1,-1,0$)  &  ($1,-1,\pm 1$)\\
\hline
$\langle H \rangle/|J|N$ & $0$   & $-1-2d/3$   & $-1-d$ \\
\hline
\end{tabular}
\end{table}

At finite temperatures the system is found to display qualitatively different behavior in different regions of the single-ion anisotropy strength. Due to the above arguments, we focus on the most interesting region of $-3/2 < d < 0$. In particular, the behaviors in a broad region of $-1.47 \lesssim d <0$ and a narrow region of $-3/2 < d \lesssim -1.47 $ are in more detail demonstrated on the selected cases of $d=-1$ and $d=-1.48$, respectively. In spite of the PSRG expectation of only one LRO-P phase transition, in the former case, there are two anomalies in the temperature variations of various thermodynamic quantities, such as the staggered magnetization, the staggered susceptibility and the specific heat, suggesting the existence of two phase transitions. From the staggered magnetization dependence in Fig.~\ref{fig:ms-T} we can observe that as the temperature is lowered some ordering is initiated already above $k_BT_{2}/|J| \approx 0.5$. The phase just below $k_BT_{2}/|J|$ is characterized by a finite value of $m_s$ for finite $L$ but there is another anomalous increase at $k_BT_{1}/|J| \approx 0.4$, which eventually leads to the expected AF LRO phase of the type $(1,-1,0)$ below the temperature $k_BT_{1}/|J|$. 

\begin{figure}[h!]
\centering
    \includegraphics[scale=0.5]{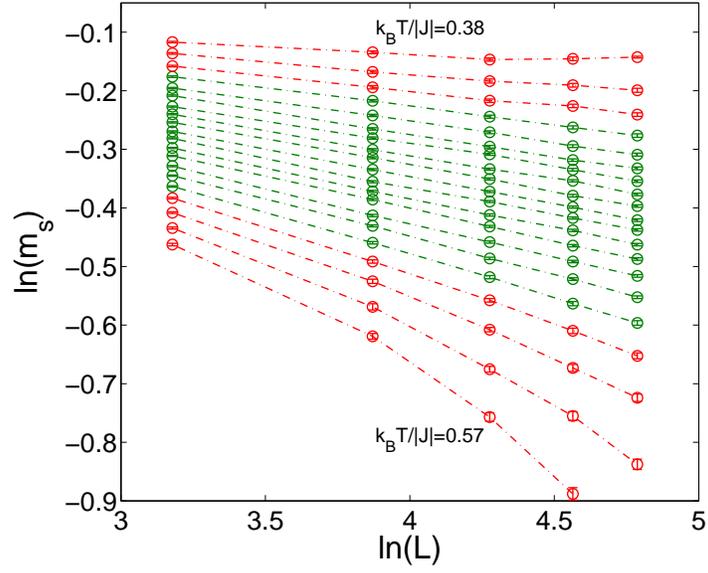}
\caption{(Color online) Log-log plots of the staggered magnetization against the lattice size for different temperatures.}\label{fig:log_ms-log_L}
\end{figure}

\begin{figure}[h!]
\centering
    \includegraphics[scale=0.5]{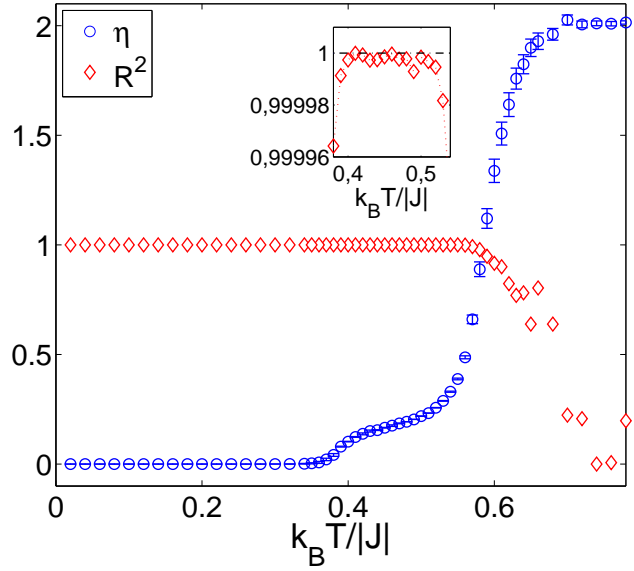}
\caption{(Color online) Temperature variation of the exponent $\eta$ and the coefficient of determination of the linear fit $R^2$ (measure of goodness of fit) obtained from the FSS analysis, for $d=-1$. The symbols and the error bars respectively represent the mean and the extreme values obtained from Eqs.~(\ref{ms_FSS}) and~(\ref{xis_FSS}).}\label{fig:eta-T_D-1}
\end{figure}

In the intermediate region between $k_BT_{1}/|J|$ and $k_BT_{2}/|J|$ the order parameter $m_s$ shows only slow decrease with the lattice size $L$. In Fig.~\ref{fig:log_ms-log_L} we present the FSS analyses of $m_s$ in the temperature region comprising the range $(k_BT_{1}/|J|,k_BT_{2}/|J|)$. Below $k_BT_{1}/|J| \approx 0.4$ the curves turn upward, which indicates that the values remain finite for $L \rightarrow \infty$, i.e., the LRO phase. On the other hand, above $k_BT_{1}/|J| \approx 0.5$ the curves turn downward, which indicates no ordering in the infinite lattice size limit. Excellent linear fits are obtained within the temperatures $0.4 \lesssim k_BT/|J| \lesssim 0.5$, indicating the power-law behavior expressed by Eq.~(\ref{ms_FSS}). Apparently, the slope and therefore also the value of $\eta$ varies with temperature. In Fig.~\ref{fig:eta-T_D-1} we plot the value of $\eta$, as well as the coefficient of determination $R^2$ as a measure of goodness of the linear fit, as functions of temperature using both Eqs.~(\ref{ms_FSS}) and~(\ref{xis_FSS}). At low temperatures $m_s$ is independent of $L$ and thus $\eta=0$. Within the interval $0.4 \lesssim k_BT/|J| \lesssim 0.5$ the value of $\eta$ varies from $0.103 \pm 0.001$ at $k_BT/|J| = 0.4$ to $0.219 \pm 0.001$ at $k_BT/|J| = 0.5$. We note that although the coefficient $R^2$ seems to be constantly equal to one up to almost $k_BT/|J| \approx 0.57$, the inset shows that in fact it starts deteriorating already at $k_BT/|J| \approx 0.53$. At still higher temperatures the linear fit is not appropriate and therefore the finite-size dependences are no longer power-law.

\begin{figure}[t]
\centering
    \includegraphics[scale=0.5]{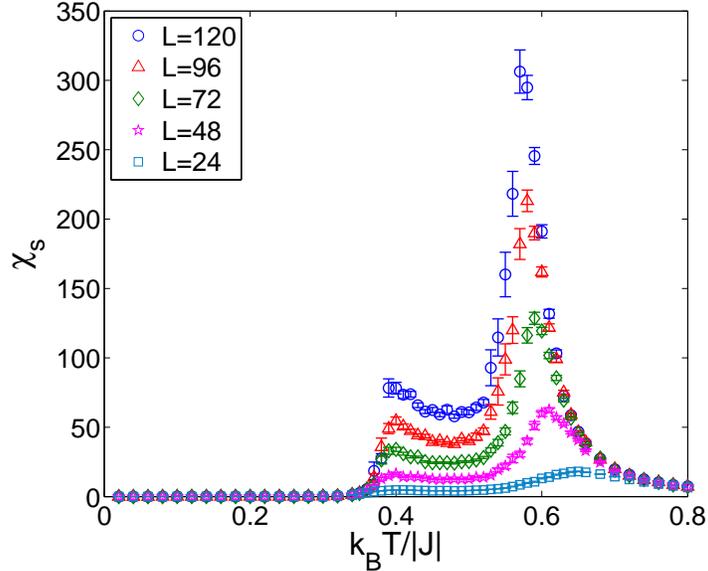}
\caption{(Color online) Temperature variation of the staggered susceptibility $\chi_s$, for $d=-1$ and different values of $L$.}\label{fig:xis-T}
\end{figure}

\begin{figure}[t]
\centering
    \includegraphics[scale=0.5]{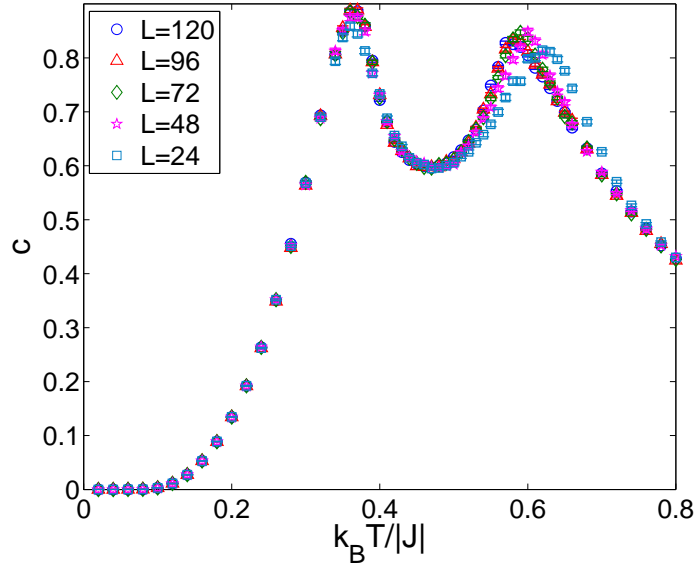}
\caption{(Color online) Temperature variation of the specific heat $c$, for $d=-1$ and different values of $L$.}\label{fig:c-T}
\end{figure}

\begin{figure}[b]
\centering
    \subfigure{\includegraphics[scale=0.4]{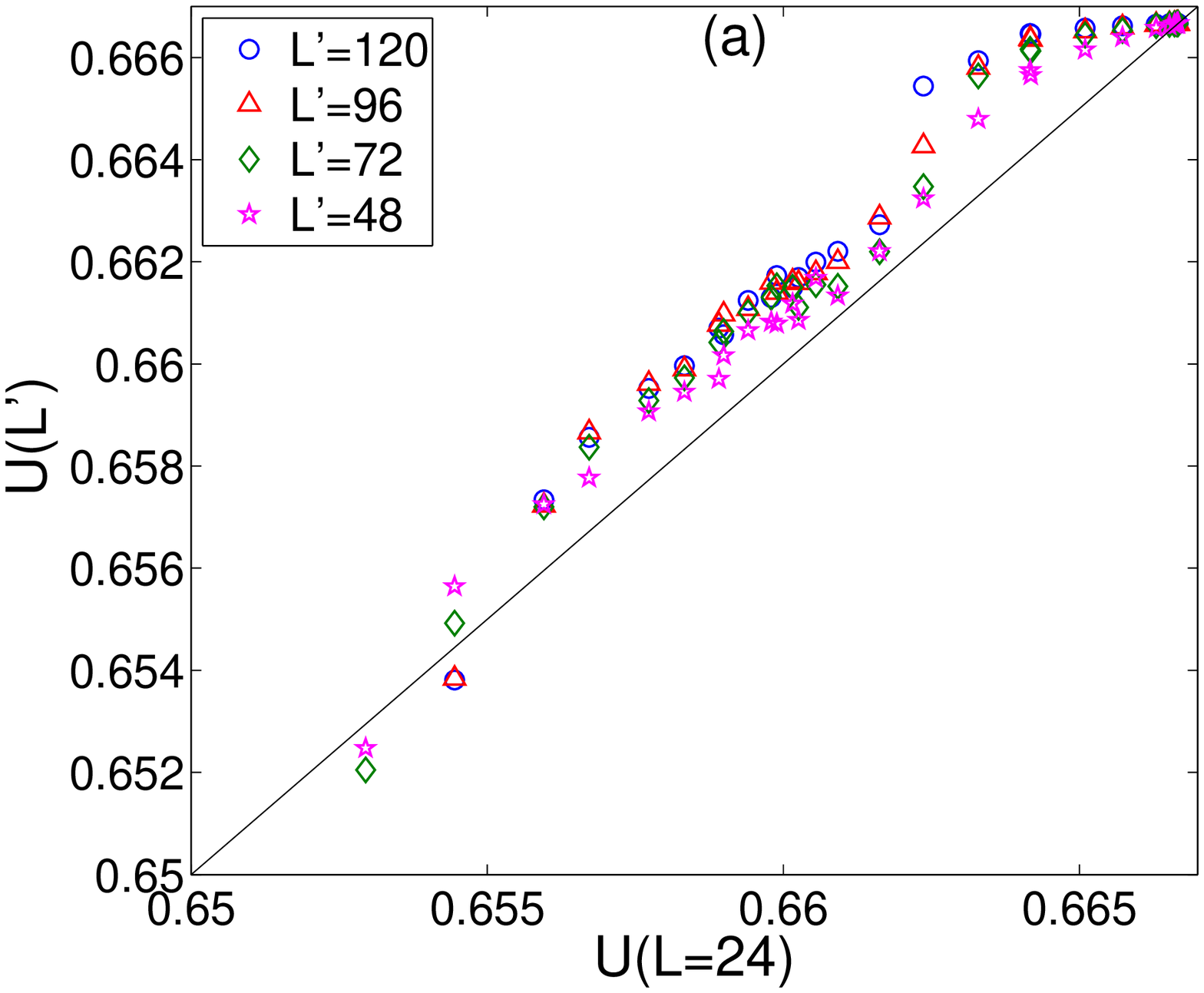}\label{fig:ms_cum_L24}}
    \subfigure{\includegraphics[scale=0.4]{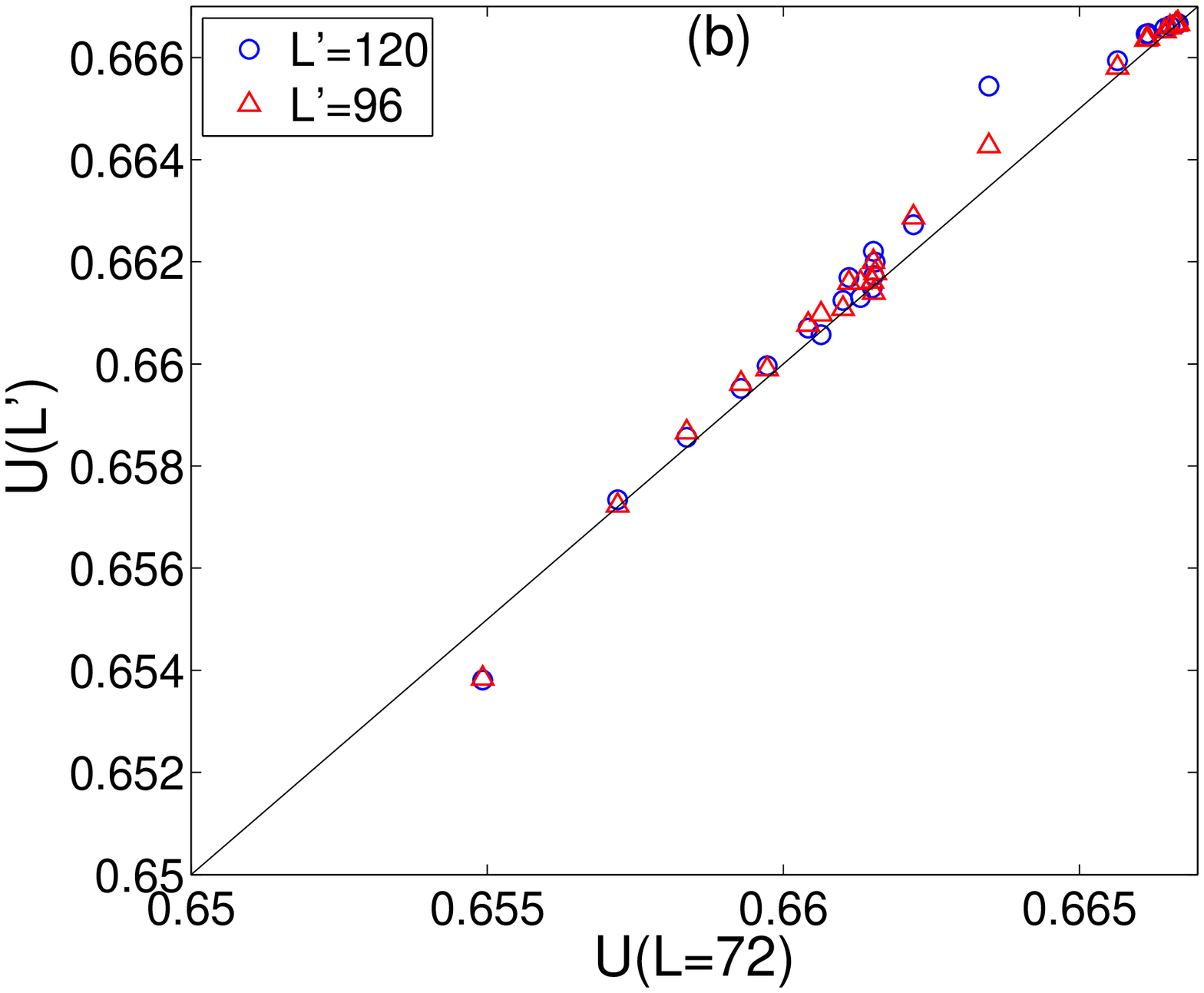}\label{fig:ms_cum_L72}}
\caption{(Color online) Binder parameter $U(L')$ plotted against (a) $U(L=24)$ and (b) $U(L=72)$ for different values of $L'>L$.}
\end{figure}

This behavior of the order parameter, along with the observations that the staggered susceptibility (see Fig.~\ref{fig:xis-T}) diverges for $L \rightarrow \infty$  and the peaks of the specific heat (Fig.~\ref{fig:c-T}) are rounded and almost independent of the lattice size above $L \approx 48$, is very similar to that observed in the TLIA model with the ferromagnetic NNN coupling~\cite{land}, the triangular lattice planar rotator model in a six-fold symmetry-breaking field~\cite{rast1}, and the six-state clock model~\cite{chal}, all displaying a BKT type of the intermediate phase. The existence of the BKT phase can be verified by analyzing the behavior of the Binder parameter $U$ according to Eq.~(\ref{U_FSS}). If the intermediate BKT phase exists it should appear in the plot as a line of points at which $U(L)=U(L')$. In Figs.~\ref{fig:ms_cum_L24} and~\ref{fig:ms_cum_L72} the parameter $U$ obtained for $L=24$ and $L=72$, respectively, is plotted against those for $L'>L$. The intersection of the resulting curve with the straight line $U(L)=U(L')$ represents a nontrivial fixed point. The hump at the values corresponding to the intermediate temperatures observed in Fig.~\ref{fig:ms_cum_L24} is apparently a finite-size effect since it disappears when large enough lattice sizes, such as those in Fig.~\ref{fig:ms_cum_L72}, are considered. Then the data for a rage of intermediate temperatures lie on a straight line $U(L)=U(L')$, implying from Eq.~(\ref{U_FSS}) that $\nu = \infty$ and, therefore, the exponential divergence of the correlation length characteristic for the BKT phase.  

\begin{figure}[t]
\centering
    \includegraphics[scale=0.5]{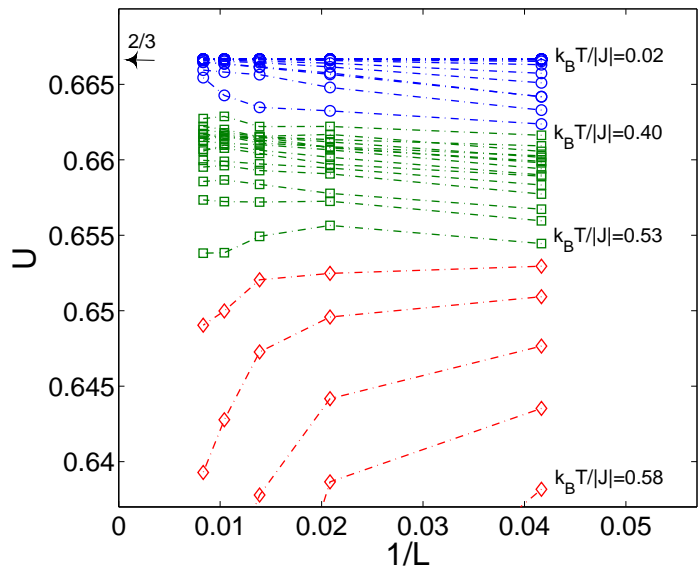}
\caption{(Color online) Lattice-size dependence of the Binder parameter $U$ for different temperatures. Circles, squares and diamonds represent typical behaviors in the LRO, BKT and paramagnetic phases, respectively.}\label{fig:U-L}
\end{figure}

The lattice-size independence of the Binder parameter $U$ in the intermediate temperature phase is further manifested in the flow diagram of $U(L)$ versus $L^{-1}$ in Fig.~\ref{fig:U-L}. If $L \rightarrow \infty$, for $k_BT/|J|<0.40$ the value of $U=2/3$ represents the trivial fixed point of the LRO phase, for $k_BT/|J|>0.53$ the value of $U=0$ represents the high-temperature fixed point of the disordered phase, and in the intermediate-temperature range for sufficiently large sizes the value of $U$ remains constant at a given temperature.

\begin{figure}[t]
\centering
    \subfigure{\includegraphics[scale=0.4]{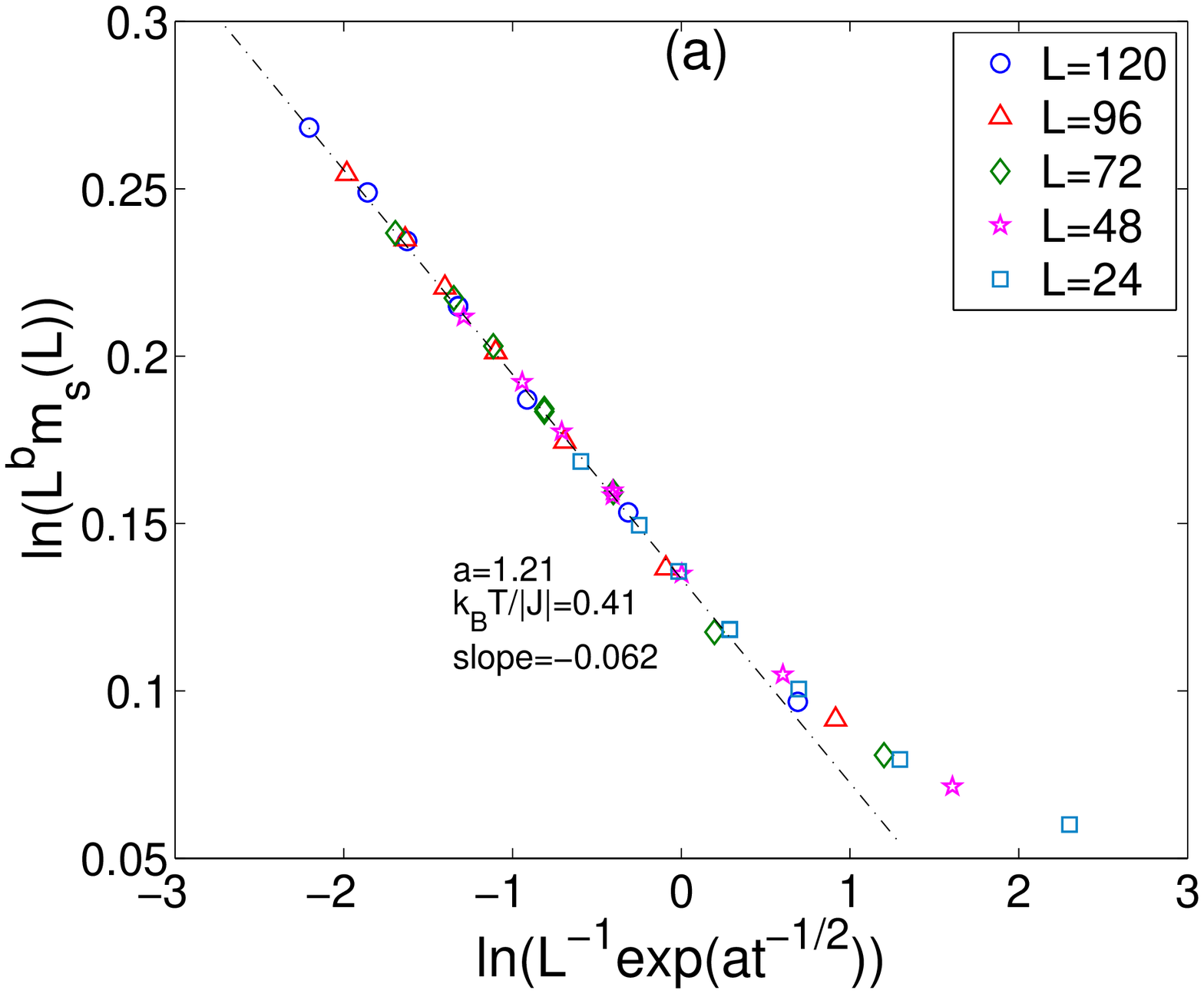}\label{fig:bkt_fss_ms}}
    \subfigure{\includegraphics[scale=0.4]{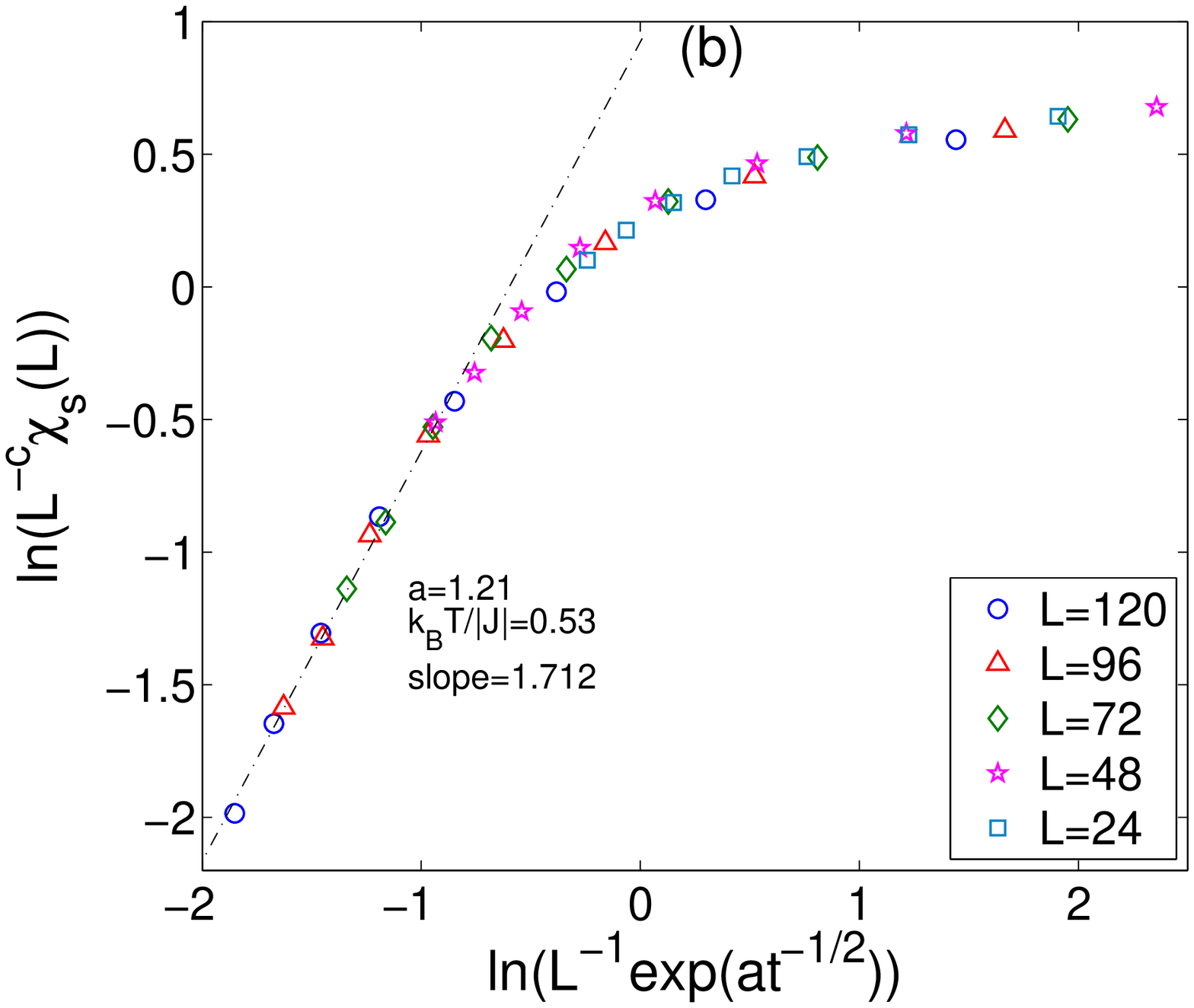}\label{fig:bkt_fss_xis}}
\caption{(Color online) Finite-size scaling of (a) the staggered magnetization $m_s$ and (b) the staggered susceptibility, according to the scaling relations~(\ref{ms_bkt}) and~(\ref{xis_bkt}), respectively.}
\end{figure}

In order to determine more precisely the values of the LRO-BKT and BKT-P transition temperatures, $k_BT_{1}/|J|$ and $k_BT_{2}/|J|$, respectively, we further employ scaling relations~(\ref{ms_bkt}) and~(\ref{xis_bkt}). In particular, we use the values of $\eta$ determined by FSS above and tune a common value of the parameter $a$ and the transition temperatures $k_BT_{1}/|J|$, $k_BT_{2}/|J|$ such a way that the log-log plots of Eqs.~(\ref{ms_bkt}) and~(\ref{xis_bkt}) collapse on the universal curves $f_1$ and $f_2$, respectively. The plots are shown in Figs.~\ref{fig:bkt_fss_ms} and~\ref{fig:bkt_fss_xis}. The best fit corresponds to the values of $a=1.21$, $k_BT_{1}/|J|=0.41 \pm 0.01$ and $k_BT_{2}/|J|=0.53 \pm 0.01$. From the slopes of the universal curves $f_1$ and $f_2$, $-b=-0.06$ and $c=1.71$, respectively, we extract the values of $\eta=0.12  \pm 0.02$ at $k_BT_{1}/|J|$ and $\eta=0.29  \pm 0.04$ at $k_BT_{2}/|J|$, consistent with the values obtained from Eqs.~(\ref{ms_FSS}) and~(\ref{xis_FSS}). We would like to point out that these values are in   a fair agreement with the $\eta$ values corresponding to the respective lower and upper limits of the BKT transition temperatures in the spin-1/2 TLIA model with competing NNN interactions~\cite{land}, the planar rotator model with six-fold symmetry breaking fields~\cite{jose,rast1} as well as the six-state clock model with both non-frustrated ferromagnetic interactions on a square lattice~\cite{chal} and frustrated antiferromagnetic interactions on a triangular lattice~\cite{suru}. The theoretical prediction for the latter are $\eta(T_1)=1/9$ and $\eta(T_2)=1/4$.

%

\begin{figure}[t]
\centering
    \includegraphics[scale=0.5]{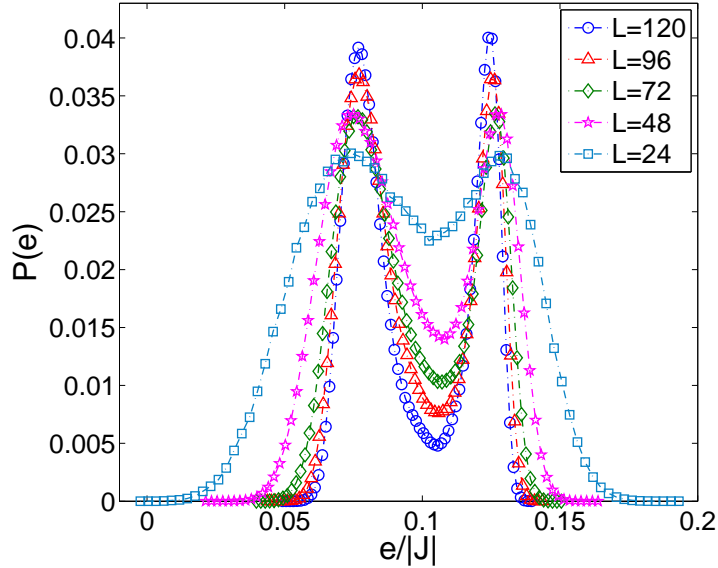}
\caption{(Color online) Energy distributions at the $L$-dependent pseudo-critical temperatures $k_BT_{c}(L)/|J|$ for $d=-1.48$. Double-peaked structure with deepening barrier between the two energy states with the increasing $L$ signals a first-order transition.}\label{fig:first_order}
\end{figure}

\begin{figure}[t]
\centering
    \subfigure{\includegraphics[scale=0.4]{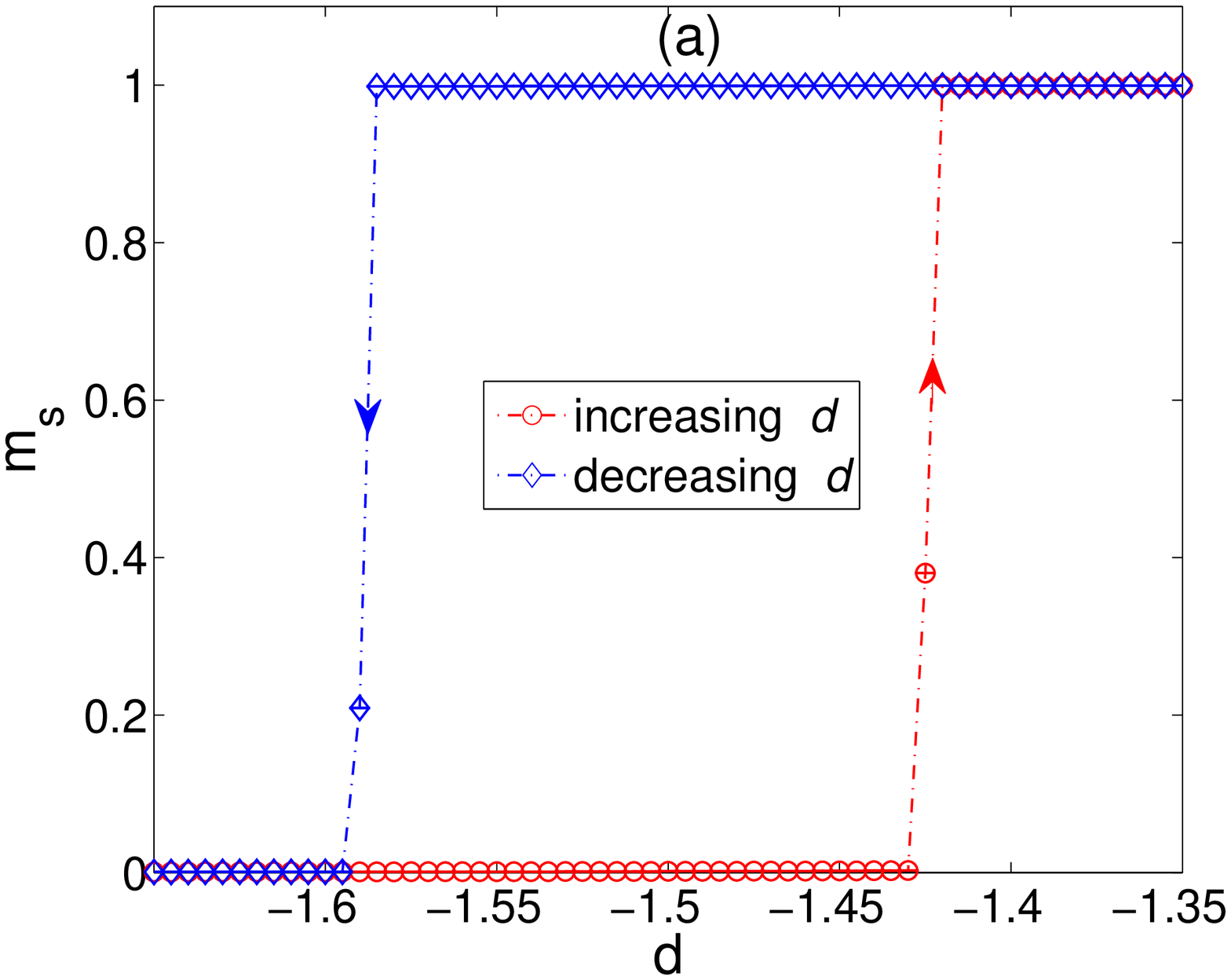} \label{fig:ms-D}}
    \subfigure{\includegraphics[scale=0.4]{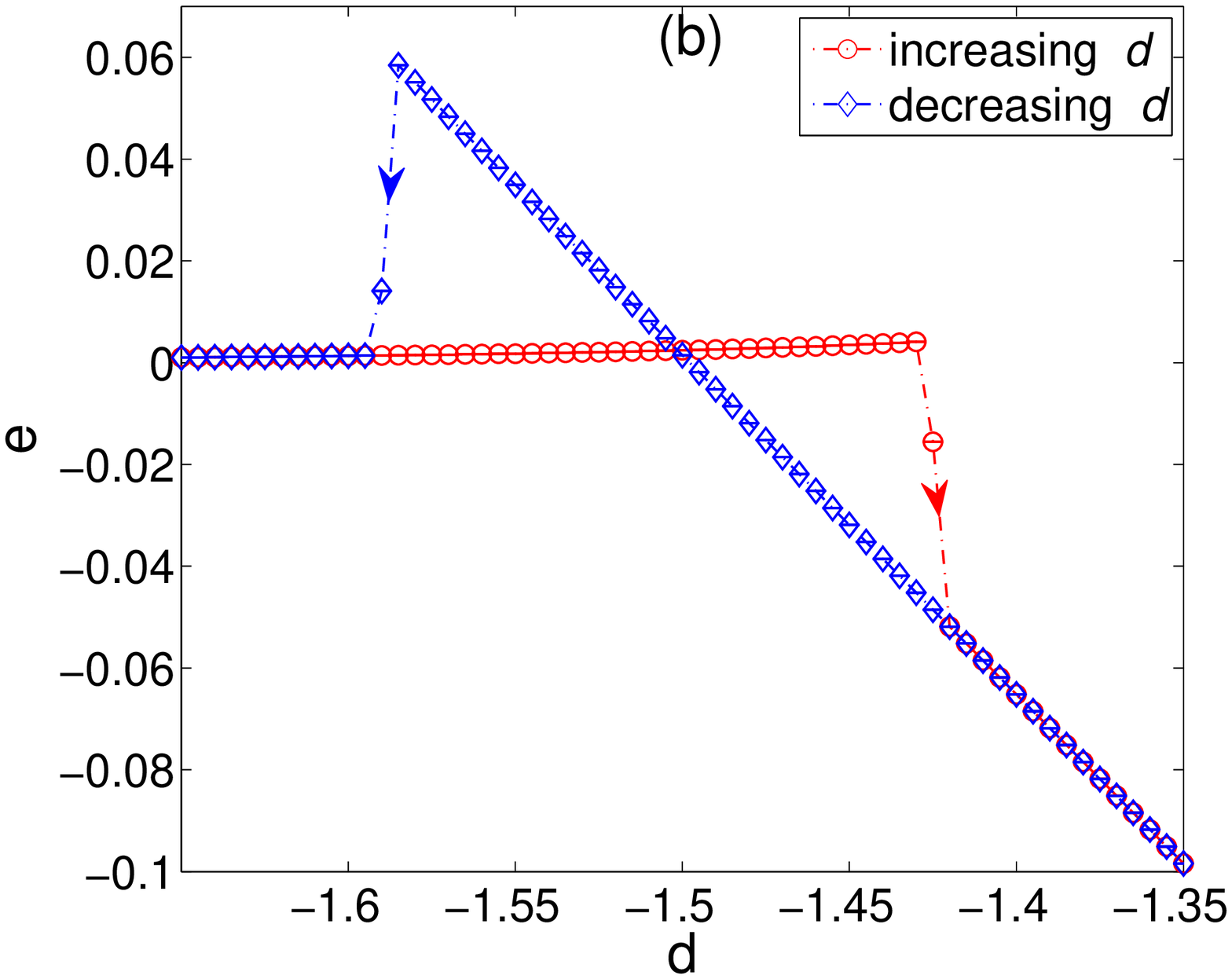} \label{fig:e-D}}
\caption{(Color online) Variations of (a) the staggered magnetization $m_s$ and (b) the internal energy $e$ in the decreasing (diamonds) and increasing (circles) single-ion anisotropy $d$ at $k_BT/|J|=0.2$.}\label{fig:hysteresis}
\end{figure}

The LRO-BKT and BKT-P phase boundaries merge at $d \approx -1.47$, below which the transition changes to the LRO-P type, i.e., between the long-range ordered and the paramagnetic phases, and the transition is of first order. The discontinuous character is evident from the energy histograms, shown in Fig.~\ref{fig:first_order} for $d = -1.48$. The histograms are bimodal with the dip between the peaks observable already at moderate values of $L$ and approaching zero as $L$ is increased. Thus our Monte Carlo estimate of the point at which the BKT transition lines merge and change to the first-order one $(d_t,k_BT_t/|J|) \approx (-1.47,0.35)$ gives the values higher than those for the presumed tricritical point obtained by the PSRG method $(d_t,k_BT_t/|J|) = (-1.494,0.300)$~\cite{maha}. Below $d = -1.48$ the transition temperature drops sharply and the tunneling times between the two modes increase enormously. At the same time the thermodynamic quantities, such as the staggered magnetization $m_s$ and the internal energy $e$, shown in Fig.~\ref{fig:hysteresis}, start displaying strong hysteretic behavior, associated with formation of metastable states, when $d$ is increased and decreased. At sufficiently low temperatures the transition point can be approximately located as an intersection point of the internal energy dependences in the $d$-increasing and $d$-decreasing processes. As evidenced from Fig.~\ref{fig:e-D}, for $k_BT/|J|=0.2$ the transition occurs at $d_c \approx -1.5$, in line with our expectations from the ground-state arguments.  

\begin{figure}[b]
\centering
    \includegraphics[scale=0.5]{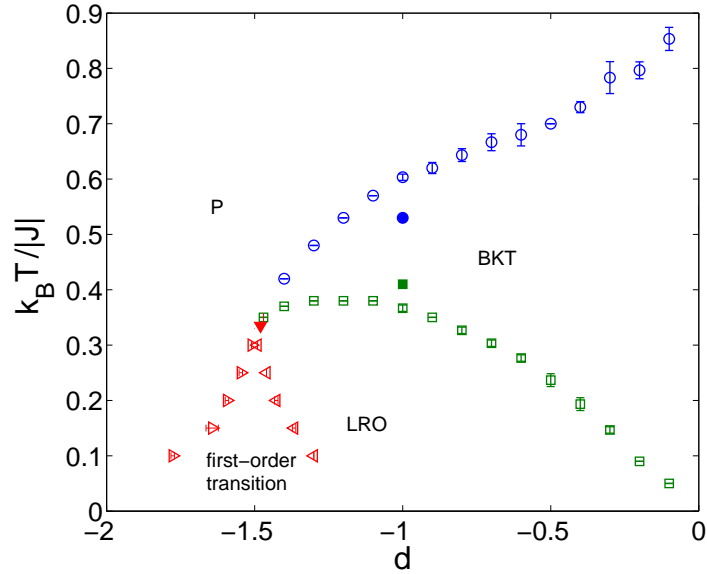}
\caption{(Color online) Rough estimate of the phase diagram in the $d-k_BT/|J|$ plane obtained from the locations of the specific heat maxima (empty symbols), including a few more precise data from the FSS analyses (filled symbols). }\label{fig:PD}
\end{figure}

\begin{figure}[t]
\centering
    \subfigure{\includegraphics[scale=0.42]{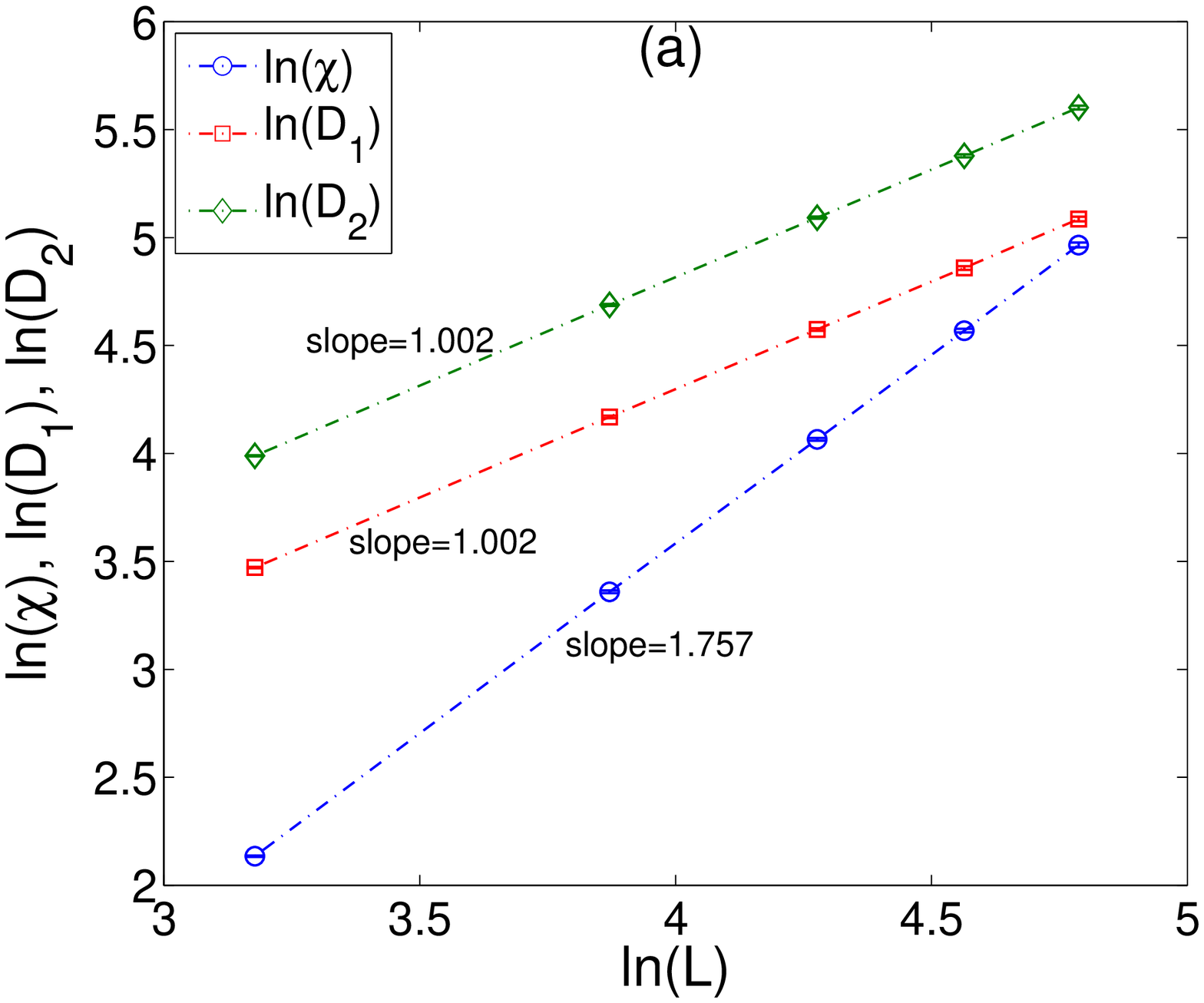} \label{fig:fss_ferro_ind}}
    \subfigure{\includegraphics[scale=0.42]{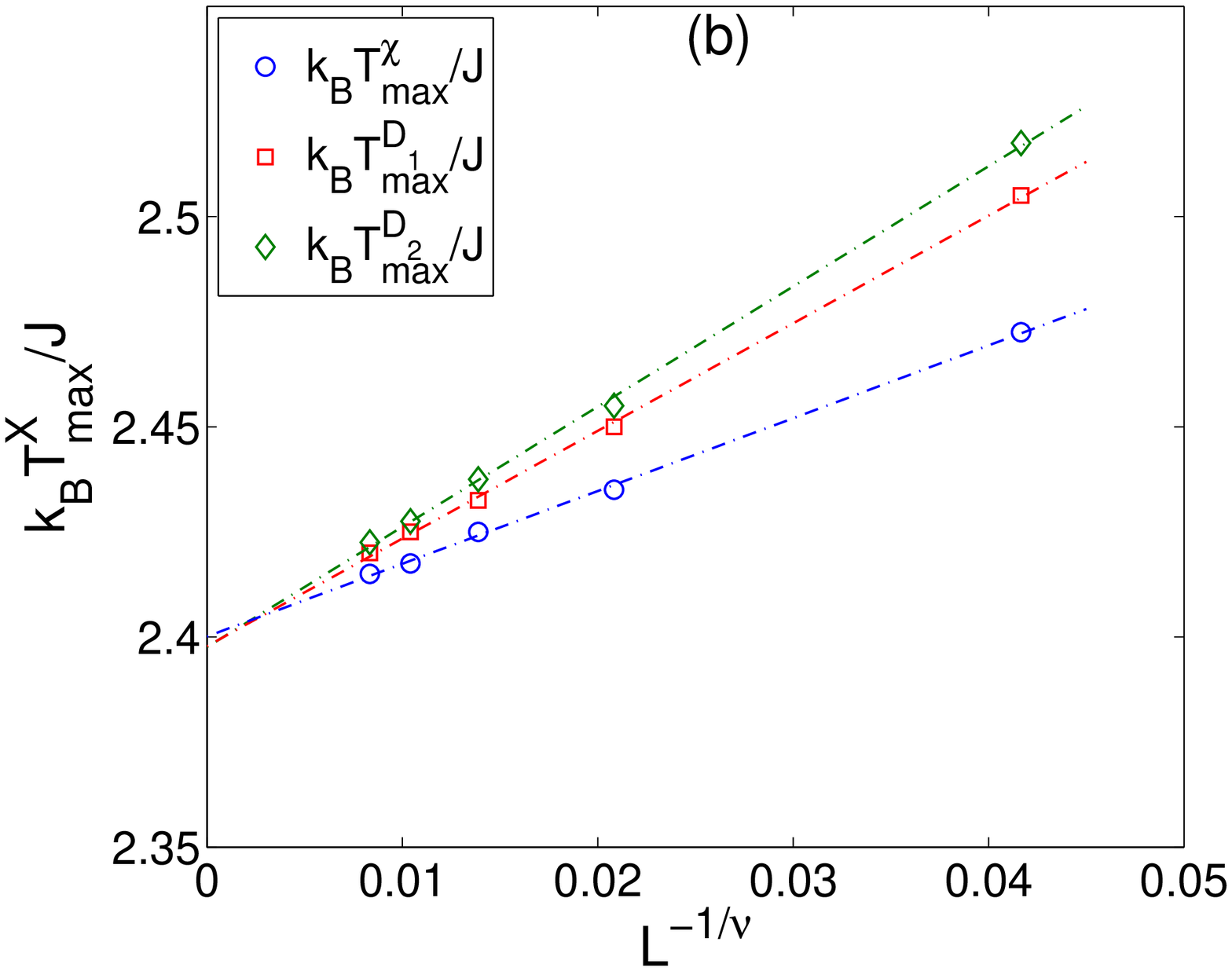} \label{fig:fss_ferro_Tc}}
\caption{(Color online) (a) FSS behavior of the maxima of the direct susceptibility $\chi$ and the logarithmic derivatives of the first and second moments of the magnetization $D_1$ and $D_2$, respectively, in a log-log plot, and (b) FSS fits of the pseudo-transition temperatures $k_BT_{max}^{X}/J$, where $X=\chi,D_1,D_2$ and $\nu=0.998$, for $J>0$ and $D/J=-1$.}\label{fig:fss_ferro}
\end{figure}

Finally, in Fig.~\ref{fig:PD} we provide a rough estimate of the phase diagram in the $d-k_BT/|J|$ plane. The boundaries denoted by the empty symbols are obtained from the specific heat maxima, using $L=48$, $N=2 \times 10^5$ and three independent MC runs. The low-temperature branches denoted by the left- and right-pointing triangles represent the jumps of the energy (and other quantities) in the $d$-increasing and $d$-decreasing measurements, respectively, and outline the two-phase coexistence region characteristic for first-order transitions below $d \approx -1.47$ (see Fig.~\ref{fig:hysteresis}). Based on the ground-state considerations, the true first-order phase transition boundary is expected to drop to $d = -1.5$ at zero temperature. The filled symbols represent the transition points determined by the FSS analyses above. It is apparent that the locations of the specific heat maxima underestimate the LRO-BKT transition temperature but overestimate the BKT-P transition temperature. This behavior is typical also for some other systems displaying the intermediate BKT phase~\cite{land,chal,rast1}. 

For the sake of comparison, we also checked the critical behavior of the same model but with the ferromagnetic interaction $J>0$ and $d=-1$. In this case we only observed one anomaly in various thermodynamic quantities, associated with the ferromagnetic-paramagnetic phase transition with the power-law scaling of various thermodynamic functions. The FSS analysis of the direct susceptibility $\chi$ and the logarithmic derivatives of the first and second moments of the magnetization $D_1$ and $D_2$\footnote{The quantities $\chi$, $D_1$ and $D_2$ are obtained from the Eqs.~(\ref{eq.chi}-\ref{eq.D2}) by replacing the staggered magnetization $m_s$ by the direct magnetization $m=(\sum_{i=1}^{L^2}S_{i})/L^2$ and the corresponding critical exponents by applying the scaling relations~(\ref{eq.scalchi}-\ref{eq.scalD2}) to these quantities.}, shown in Fig.~\ref{fig:fss_ferro_ind}, indicate that the transition belongs to the standard Ising universality class with the critical exponents $\nu_I=1$ and $\gamma_I=1.75$. Both the critical exponents $\nu=0.998 \pm 0.011$, $\gamma=1.754 \pm 0.014$ and the transition temperature $k_BT_c/J=2.399 \pm 0.002$, estimated from the scaling relation $k_BT_{max}^{X}/J=k_BT_{c}/J+aL^{-1/\nu}$, where $k_BT_{max}^{X}/J$ is the temperature at which the quantity $X$ displays a maximum and $\nu$ is the critical exponent estimated above (see Fig.~\ref{fig:fss_ferro_ind}), are in a good agreement with the recent high-accuracy MC study results~\cite{fyta}.

\section{CONCLUSIONS}
We have studied the critical behavior of the BC antiferromagnet on a triangular lattice by Monte Carlo simulations and found two kinds of phases within the single-ion anisotropy strength $-1.47 \lesssim d < 0$. Below $k_BT_{1}/|J|$ the system displays the antiferromagnetic LRO on two sublattices with the third one remaining in a non-magnetic state. Above $k_BT_{1}/|J|$ for a range of temperatures up to $k_BT_{2}/|J|$, the ordering is of the BKT-type with a power-law decaying spin-correlation function. For $-1.5 \le d \lesssim -1.47$, there is only one phase transition from the LRO to the paramagnetic region and the transition is of first order. This behavior is distinctively different from both the ferromagnetic BC model on a triangular lattice and also from a non-frustrated antiferromagnetic model on a bipartite lattice~\cite{kimel}, which do not exhibit the BKT phase and the second-order transition from the LRO to the paramagnetic phase is of the standard Ising universality class. However, there are some other systems, such as a $q$-state clock model with $q>4$, which have been confirmed to display similar critical behavior to the present model, featuring the low-temperature LRO, the intermediate-temperature BKT and the high-temperature paramagnetic phases. Furthermore, for a selected value of the single-ion anisotropy $d = -1$ the current BC model produced the BKT phase with the values of the temperature-dependent exponent $\eta$ in the high- and low-temperature limits consistent with the theoretical predictions for the planar rotator model with six-fold symmetry breaking fields, $\eta(T_2)=1/4$ and $\eta(T_2)=1/9$~\cite{jose}, as well as those estimated by Monte Carlo simulations in the spin-1/2 TLIA model with the ratio of competing NN and NNN interactions equal to one~\cite{land}, the planar rotator model with six-fold symmetry breaking fields~\cite{rast1} as well as the six-state clock model~\cite{chal,suru}. All these models share the six-fold ground-state degeneracy, which we believe is behind the universal behavior at finite temperatures.


Further, it would be interesting to see how the phase diagram evolves if larger spin values are considered. Based on the earlier studies~\cite{yama}, for $S$ larger than some critical value $S_c$ in the ground state the LRO should set-in already at $d=0$ with the BKT phase transitions still occurring at higher temperatures. On the other hand, the increasing spin number is believed to change the multicritical behavior in the large $|d|$ limit of the BC model~\cite{xavi,ping}. Thus, our future intention is to extend the present investigations to the systems with larger spin values and focus on peculiarities arising from the presence of the geometrical frustration.

\begin{acknowledgments}
This work was supported by the Scientific Grant Agency of Ministry of Education of Slovak Republic (Grant No. 1/0234/12). The authors acknowledge the financial support by the ERDF EU (European Union European regional development fund) grant provided under the contract No. ITMS26220120005 (activity 3.2).
\end{acknowledgments}

\end{document}